# The anatomy of conspiracy theorists: Unveiling traits using a comprehensive Twitter dataset

Margherita Gambini [a,b,1], Serena Tardelli [b,*,1], Maurizio Tesconi [b]

[a] Department of Information Engineering, University of Pisa, Via Girolamo Caruso, 16, Pisa, 56122, Italy

[b] Institute of Informatics and Telematics, National Research Council, Via Giuseppe Moruzzi, 1, Pisa, 56127, Italy

ABSTRACT

The discourse around conspiracy theories is currently thriving amidst the rampant misinformation in online environments. Research in this field has been focused on detecting conspiracy theories on social media, often relying on limited datasets. In this study, we present a novel methodology for constructing a Twitter dataset that encompasses accounts engaged in conspiracy-related activities throughout the year 2022. Our approach centers on data collection that is independent of specific conspiracy theories and information operations. Additionally, our dataset includes a control group comprising randomly selected users who can be fairly compared to the individuals involved in conspiracy activities. This comprehensive collection effort yielded a total of 15K accounts and 37M tweets extracted from their timelines. We conduct a comparative analysis of the two groups across three dimensions: topics, profiles, and behavioral characteristics. The results indicate that conspiracy and control users exhibit similarity in terms of their profile metadata characteristics. However, they diverge significantly in terms of behavior and activity, particularly regarding the discussed topics, the terminology used, and their stance on trending subjects. In addition, we find no significant disparity in the presence of bot users between the two groups. Finally, we develop a classifier to identify conspiracy users using features borrowed from bot, troll and linguistic literature. The results demonstrate a high accuracy level (with an F1 score of 0.94), enabling us to uncover the most discriminating features associated with conspiracy-related accounts.

## 1. Introduction

Conspiracy culture has been brewing on social media for over a decade, fueled by the misinformation, polarization, and science denial typical of the online ecosystem [1–4]. Conspiracy theories provide alternative explanations to significant historical or current events with claims of secret plots by people or groups having ambiguous intentions (e.g., usurpation of power, violation of rights, alteration of the bedrock institutions, societal disruption, etc.) [5–7]. Social media platforms have enabled faster communication and dissemination of conspiratorial narratives. As such, recent times have seen a plethora of online conspiracy beliefs concerning a broad range of topics. Notable examples encompass unconventional interpretations of climate change [8], the 9/11 attacks [9], political movements like QAnon [10], and, more recently, theories related to the COVID-19 pandemic [11,12]. As a result, the spread of such information can have far-reaching implications for both individual users and society at large [13–17]. For these reasons, research into online conspiracy has grown in recent years, aimed at comprehending the dynamics of online conspiracy culture across various academic disciplines by using models and analytical approaches primarily based on linguistic and rhetorical theory [7]. Understanding users' inclinations towards conspiracy theories is of significant interest, as it can offer valuable insights into the propagation of ideologies, without limiting the analysis to a specific conspiracy theory. This understanding is crucial for assessing the roles played by the involved individuals and taking appropriate measures to mitigate the impact of this phenomenon. Nonetheless, despite research advances in the detection of emerging or predefined conspiracy theories [18–20], the study of conspiracy users' characteristics remains limited. In fact, most of the existing datasets and collection techniques are either focused on specific conspiracy topics or domains, or rely on manual annotation or subjective criteria to label users as conspiratorial or not. This limitation poses challenges for developing and evaluating automated methods for user-level conspiracy detection and analysis. Therefore, there is a need for a large-scale, diverse, and reliable dataset that can capture the general characteristics and behaviors of conspiracy users across different topics. Such a dataset would also enable researchers to explore various aspects of online conspiracy, such as network analysis, content analysis, sentiment analysis, misinformation and stance detection.

* Corresponding author.
*E-mail addresses:* margherita.gambini@iit.cnr.it (M. Gambini), serena.tardelli@iit.cnr.it (S. Tardelli), maurizio.tesconi@iit.cnr.it (M. Tesconi).

[1] These authors contributed equally.

### 1.1. Contributions

In this study, we collect and share a Twitter dataset of 15K users, categorized into two groups: a set of *conspiracy users* engaging with diverse conspiracy theories in May 2022, and a control group of *random users* collected from those posting on the same topics during the same period. Additionally, we collect their timelines, totaling of 37M tweets. Numerous social platforms, ranging from fringe to mainstream, have been exposed to various types of conspiracy narratives [10,21–27]. Here, we focus on Twitter data due to its reported extent of conspiracy engagement, wide audience reach, rapid dissemination, and ease of accessibility. With a robust dataset, we analyze the distinctions between the two user groups across three dimensions: topic preferences, profile metadata, and behavioral patterns.

In particular, our approach allows us to explore new research directions and address the following research questions:

**RQ1** — *How can we construct a robust and comprehensive dataset of online conspiracy users?* Existing datasets are often limited by simplistic gathering methods. In this study, we collect with a rigorous methodology users endorsing conspiracy beliefs posted by known conspiracy source on Twitter. We also collect a control group of random users with similar metadata properties who discuss the same topics but do not show signs of conspiracy involvement, ensuring a fair comparison between the two groups.

**RQ2** — *What are the differences in attitude and approach when discussing topics between conspiracy and random users?* Different user groups can interpret and engage with topics in diverse ways. We delve into and compare the manners in which these two groups relate to and discuss specific subjects.

**RQ3** — *Which features predominantly differentiate the conspiracy users?* Our analysis of feature importance involves training various classifiers using several classes of features. As such, we show the most characteristics traits that describe conspiracy theorists. Furthermore, we compare our findings with a state-of-the-art technique and dataset that leverage deeper linguistic properties.

Our main contributions stemming from the aforementioned RQs can be summarized as follows:

- We create and publicly share a large, robust, and balanced dataset of 15K conspiracy and random control users, along with 137M tweets from their timelines.
- We show that these two user groups display divergent attitudes and perspectives on specific topics, thereby reinforcing their distinctions.
- Our analysis indicates that both groups have an automation rate below 1%, suggesting the involvement of genuine, real individuals.
- We provide analysis on discriminating features by employing a classifier that leverages profile metadata, behavioral characteristics, and linguistic features borrowed from the literature on bot, troll, and conspiracy detection, obtaining a high F1 score of 0.94.

**Reproducibility.** We release an anonymized, privacy-preserving version of the dataset.[2]

[2] https://zenodo.org/record/8239530

### 1.2. Roadmap

The remainder of this paper is organized as follows. In Section 2, we conduct a review of recent literature on online conspiracies, emphasizing works that provide accessible datasets comparing conspiracy users to random users. In Section 3, we outline our data collection strategy, offering initial descriptive statistics of the resulting dataset. In addition, we provide an overview of the features leveraged for the analysis and characterization of conspiracy users compared to a control group of random users. In Section 4, we present the results, highlighting distinctions between the two groups in terms of topic and profile metadata, as well as present the classification results to identify the most discriminating features. Lastly, in Section 5, we summarize the key findings, contextualize them within the broader research on online conspiracy, address the limitations, and suggest potential avenues for future research.

## 2. Related work

In recent years, there has been an increasing focus on identifying and understanding conspiracy theories. Unlike other forms of misleading content, like misinformation, disinformation, fake news, or rumors, conspiracy theories present a unique challenge. They raise the crucial and ongoing to question whether they are fundamentally false, alternative explanations, or situated somewhere between reality and fiction [5, 7]. Interestingly, individuals on social media who embrace conspiracy theories are not solely automated accounts, trolls, or spreaders of fake news and rumors. Many genuinely hold beliefs in hidden agendas or plots, some of which have been verified as true,[3] while others remain unverified. Furthermore, not all conspiracy theorists actively scheme or propagate them. Among them are regular people who simply hold these beliefs without actively disseminating them. To discriminate between these different types of users, the availability of a high-quality, robust social media dataset is essential. Such a dataset must encompass the wide array of dynamic behaviors associated with conspiracy beliefs, but its creation presents significant challenges that necessitate meticulous planning and assessment. Previous studies have attempted to create social datasets for similar purposes, not without some limitations and trade-offs.

The following two subsections focus on the current leading methods for constructing datasets comprising both conspiracy-affiliated users and those in a control group, as well as describe how these two groups are compared and characterized.

### 2.1. Datasets of conspiracy and control group users

In literature, the comparison between users who engage in conspiracy theories and a control group is commonly performed by analyzing the presence of specific keywords and/or URLs related to these theories within social media posts [20,21,28–33]. In this context, the control group typically consists of users who consume content that directly opposes specific conspiracy theories. Some studies employed such approach and collected datasets of conspirative and non-conspirative users. For instance, the authors in [28] collected and annotated tweets related to COVID-19 conspiracy, sampled 109 conspirative users annotated as posting conspiracy theory content and retrieved their timelines. As control group, they identified 109 non-conspirative users exhibiting a tweet behavior focusing on coronavirus related content in general, using generic related keywords (e.g., corona, covid, pandemic). However, this study faces certain limitations. Notably, the dataset size is relatively small due to the challenging task of manual annotation, and the focus is narrowed to a specific conspiracy theory.

[3] https://www.rd.com/list/conspiracy-theories-that-turned-out-to-be-true/

**Table 1**
State-of-the-art datasets for analyzing conspiracy theorist users. Studies focusing on detecting conspiracy theories, as opposed to user-centric approaches, did not establish distinct user groups (conspiracy theorists and control group), as indicated by "N/A".

| Ref. | Focus theory | Focus level | Collection strategy | Period collection | Timelines/ tweets | Conspiracy theorists | Control group |
|---|---|---|---|---|---|---|---|
| Batzdorfer et al. [28] | specific | user | keywords | 11 months | 600K | 109 | 109 |
| Giachanou et al. [29] | specific | user | keywords | 1 month | 2M | 977 | 950 |
| Fong et al. [30] | general | user | followers | 5 days | 170K | 880 | 786 |
| Pogorelov at al. [33] | specific | tweet | keywords | 15 months | 3K | N/A | N/A |
| Rains et al. [31] | specific | tweet | keywords | 3 months | 1.8M | N/A | N/A |
| Our work | general | user | likes | 7 months | 37M | 7394 | 7394 |

Similarly, prior investigations have predominantly centered around specific conspiracy theories [33–37]. In [31], the authors focused on the conspiracy narrative surrounding COVID-19, leveraging hashtags in support or opposed to specific conspiracies. Similarly, the authors in [20] collected and analyzed online discussions related to four distinct conspiracy theories. Nevertheless, findings tied exclusively to a single conspiracy theory may not be easily extrapolated to other societal events that might trigger beliefs in conspiracies. Therefore, our primary objective is to explore users involved in conspiracy-related discussions without being confined to a particular information operation or theory.

The authors in [29] advanced the state-of-the-art by considering six conspiracy theories. They curated and annotated tweets containing hashtags likely used to either endorse or refute these theories. This resulted in retrieving 977 users who engage with conspiracy content and 950 users who counteract such narratives. Meanwhile, the authors in [30] managed to expand the conspiracy user base by adopting a different method. They exploited five known conspiracy-affiliated and five science-oriented Twitter influencers, retrieving a sample of their followers to create a dataset of pro- and anti-conspiracy groups. However, these approaches have some drawbacks. Specifically, hashtags or relationships do not always accurately reflect the content of a tweet, as they can indicate both support for or rejection of conspiracy theories. Our approach tackles this issue by considering *likes*, which better capture individual user preferences. This eliminates the need for manual annotation and allows for the examination of a larger user base.

Close to our approach, previous works [21,32] labeled conspiracy enthusiasts and science-minded users based on the number of likes they gave to conspiracy and science-related posts on Facebook. However, their analysis was confined to the comments these users left in conspiracy and science groups, disregarding their broader Facebook posting history. Finally, it is worth highlighting that our focus is on comparing conspiracy theorists with the broader population of Twitter users, rather than exclusively contrasting them with those who oppose predefined conspiracies. Our focus is on Twitter, for its (formerly) data retrieval ease and its role in the dissemination of conspiracy theories [38,39]. However, the approach we employ holds the potential for application across various social media platforms. In summary, Table 1 provides a comparison between user-centric Twitter datasets available in literature, highlighting distinctions between existing datasets and the dataset we have meticulously curated and analyzed in this study.

### *2.2. Characterization of users engaging conspiracies*

The characterization and detection of malicious users has received a lot of attention during the last years. Researchers have primarily concentrated on analyzing the traits of various types of problematic users, such as social bots, state-sponsored trolls, and more recently, conspirators. These investigations have predominantly focused on elements like profile metadata, demographic details (e.g., gender, age) [40–42], social activity [43], interactions [42,44], and relationships [12,16,22, 42].

A more recent trend in conspiracy detection involves examining linguistic patterns present in textual content [22,29,30,32]. This approach aims to identify specific language cues associated with conspiratorial discourse. For instance, the authors in [43] explored linguistic features like average text length, word redundancy, emotional responses, and psycholinguistic aspects to distinguish conspiracy-related content. Likewise, other researchers explored linguistic cues to detect trolls [45] and conspiracy [12,29,46]. Similarly, our work investigates features commonly used to spot malicious users (such as fake accounts, bots, trolls, and spreaders of misinformation), which have not been extensively explored for identifying conspiracy theorists [47], while also incorporate linguistic features into our analysis.

To obtain the features that better discriminate conspirative users from random ones, we leverage the outcomes from several standard machine learning classifiers. Similar efforts were made by previous works, such as [30], in which authors employed a logistic regression model to confirm their qualitative findings on psycholinguistic traits associated with conspiracy believers and science enthusiasts. However they did not present detection results for direct comparison with our work. In another relevant study, [33], the authors analyzed covid19 conspiracy discussions and classified users into misinformation and non-misinformation groups based on their profile metadata and tweet embeddings. However, their approach relied on graph-based methods and did not provide sufficient details for reproducibility (e.g., dataset, graph-data), which limits the comparison with our work.

In contrast, our study proposes a model that encompasses diverse features to identify online users engaged in conspiracy discussions. We also benchmark our classification outcomes against a study, [29], that used a CNN-based model incorporating linguistic traits to differentiate between users who share posts supporting or refuting conspiracy theories.

In summary, we adopt a computational approach to study the conspiracy phenomenon and compare online users who engage in conspirative discussions with random users. In particular, we examine the profile, activity and psycholinguistic characteristics of conspiracy and random users based on the tweets that they post. Furthermore, we introduce a model that utilizes different features to identify online users participating in conspiracy-related conversations. Therefore, our study offers an orthogonal view and contribution building on prior work by employing distinct methodologies for constructing a reliable and robust conspiracy dataset and applying extended methodologies for characterization and detection.

## 3. Methods and data

In this section, we delve into the methods employed to gather and analyze data for our study. We first delineate our data collection strategies for collecting users engaged in conspiracy theories and random users. Next, we describe feature extraction methodologies.

### *3.1. Data collection strategy*

To address our first research question (RQ1) on creating a robust dataset of users engaged in conspiracy content, we propose a strategy based on users' *liking* behavior towards posts from various conspiracy accounts. We argue that liking a post indicates stronger approval and endorsement of the message compared to a mere re-share [48].

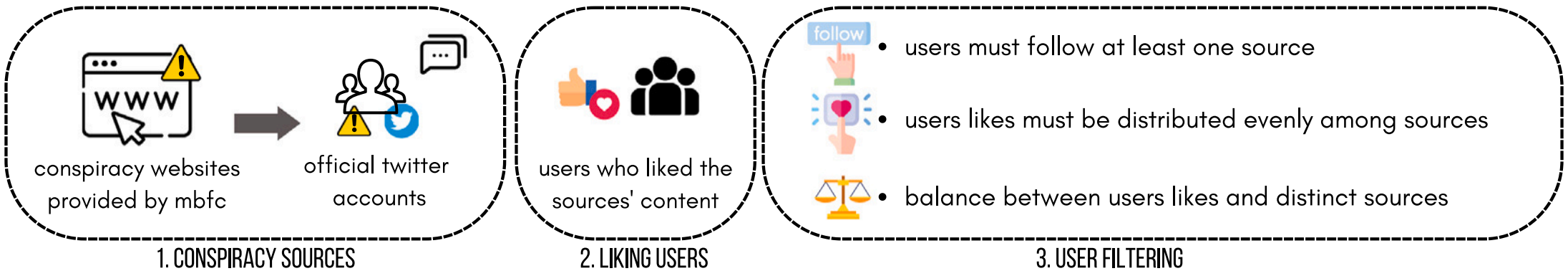

**Fig. 1.** Overview of the proposed strategy for collecting conspiracy theorist users.

To ensure a fair comparison, we introduce a control group, on the line of prior research [22]. This control group comprises randomly selected users with similar metadata and engaging in discussions around the main topics. In this way, we can compare the two groups fairly.

In summary, our dataset consists of a conspiracy group comprising 7394 Twitter users and a control group with an equal number of randomly selected users. We will refer to the control group users as *random users* throughout this work. The subsequent sections describe in detail the strategy and criteria used to select both conspiracy and random users.

### *3.1.1. Strategy for collecting conspirative users*

The idea behind the collection is to identify the users who are most likely to believe in various conspiracy theories, by focusing on those appreciating, by means of a like, posts from various conspiracy sources (e.g., websites). The adopted strategy took place in June 2022 and led to the collection of the latest 3200 tweets for 7394 users known for liking conspiratorial content. Our approach comprises three key steps, which correspond to Fig. 1:

1. **Step 1 — Initial Set of Conspiracy Sources**: We identify an initial set of websites rated as conspiracy Media Bias/Fact Check (MBFC),[4] a non-profit organization assessing online source credibility and bias. We extract the associated Twitter accounts and leverage them as *seed accounts* for the following steps.
2. **Step 2 — Collecting User Likes**: We gather Twitter users who have liked posts from the seed accounts, indicating potential conspiratorial engagement.
3. **Step 3 — Applying Filters**: We retain users who both follow at least one seed account and have shown uniform interest across multiple seed accounts. The goal is to balance between diversity (number of distinct seed accounts liked by a user) and intensity (total user likes across seed accounts). Additionally, we strike a balance between the absolute number of likes per user for seed accounts and the number of liked seed accounts. We retain users based on this trade-off, constituting our conspiracy group.

In summary, our goal is to identify users likely to be a conspiracy theorist based on interactions with seed accounts. We apply the method as follows:

***Step 1****.* We select an initial set of 26 conspiracy sources from MBFC as seeds, as shown in Table 2. As mentioned, MBFC is an independent website that aims to provide an objective and transparent assessment of the credibility and bias of online news media sources. By rating more than 4K media sources and employing a team of trained experts and journalists across the political spectrum, MBFC has become the most comprehensive media bias resource on the internet. The website's ratings are based on rigorous criteria and methodology, and have been utilized by researchers for various academic purposes [49,50]. In particular, we leverage a list of 300 news source websites rated as engaging with various topics of dubious veracity and scientific validity, conspiracy theories and pseudo-science. We then look for the Twitter accounts associated with these websites and find 100 matches. For computational time purposes, we apply manual annotation and filter the matches to keep only the top 26 accounts that have more than 90 tweets endorsing any conspiracy theory in their most recent 100 tweets. Finally, we leverage them as the seed set of conspirative Twitter seed accounts.

***Step 2****.* In this step, we leverage the like interaction on Twitter, which allows users to express their appreciation or interest in a tweet. Unlike the sharing, which amplifies the message to a wider audience and allow for fact-checking or criticism, or the replying, which initiates a conversation or provides feedback that may or may not agree with the message, the like interactions convey approval or endorsement with posts.

By employing Twitter API V2 endpoints, we retrieve likes from seed accounts' posts between July 19th, 2021, and February 28th, 2022. This method captures user engagement with seed accounts over time, providing a comprehensive view. We obtain $8,935,961$ likes from 968,824 users for 54,559 tweets by seed accounts.

***Step 3****.* In this step, we perform a series of filtering steps to refine our user selection process. Initially, out of a pool of 968,824 potential conspiracy theorists, we retained only those (378,144 users) who also follow at least one of the seed accounts. This demonstrated not only an initial attraction to the tweets of these seed accounts but also an ongoing interest for their overall content. Subsequently, we filter users (345,936 users) who display a well-balanced engagement with multiple seed accounts. In other words, if a user likes content from several seed accounts, their level of interest across these accounts is relatively consistent. We measured this by applying a coefficient of variation ($Cov$) of the number of likes per seed account, keeping it at or below $1$. Finally, as mentioned, we establish the third filter based on two key factors: the total number of likes given to the set of seed accounts and the count of liked seed accounts. These factors provide insight into the intensity of activity related to conspiracy sources and the range of interest in different conspiracy theories. The combined analysis of these factors is summarized in Table 3. The table cross-references the absolute number of likes (up to 35) along the $X$-axis with the number of distinct seed accounts liked (up to 7) along the $Y$-axis. Each cell in the table represents the count of users who have distributed a minimum of Y likes across a minimum of X distinct seed accounts.

As we move towards the lower right corner of the table, the number of users naturally decreases. Ideally, we would like to select conspiracy theorists who extensively liked a large number of seed accounts. However, this approach would yield a very small pool of users, potentially less than $100$. Instead, our strategy is to strike a balance by focusing on a diverse range of sources while maintaining a reasonable number of likes. The aim was to have around 10,000 users for meaningful analysis. This process culminated in selecting 7394 conspiracy users for in-depth analysis. These users liked at least 4 different conspiracy sources and exhibited a consistent distribution of at least 25 likes.

By leveraging Twitter API, we gathered the timeline (the most recent 3200 tweets) from these 7394 selected conspiracy theorists, accumulating a total of $18,273,565$ tweets. The final dataset comprises tweets covering the time span from February 27th, 2008 to June 13th, 2022.

[4] https://mediabiasfactcheck.com/

**Table 2**
The 26 selected conspiracy websites.

| Conspiracy source provided by MBFC | Source website | Twitter account |
|---|---|---|
| Disclose TV | disclose.tv | @disclosetv |
| amtv | amtvmedia.com | @amtvmedia |
| Daily Grail | dailygrail.com | @DailyGrail |
| Catholic | catholic.org | @CatholicOnline |
| Dark Journalist | darkjournalist.com | @darkjournalist |
| End Time Headlines | endtimeheadlines.org | @EndTimeHeadline |
| CLG News | legitgov.org | @legitgov |
| Friends Of Science | friendsofscience.org | @FriendsOScience |
| Coast To Coast Am | coasttocoastam.com | @coasttocoastam |
| GeoEngineering Watch | geoengineeringwatch.org | @RealGeoEngWatch |
| Blacklisted News | blacklistednews.com | @BlacklistedNews |
| CharismaNews | charismanews.com | @charisma_news |
| Children's Health Defense | childrenshealthdefense.org | @ChildrensHD |
| GMWatch | gmwatch.org | @GMWatch |
| Creation | creation.com | @creationnews |
| Food Babe | foodbabe.com | @thefoodbabe |
| Architects and Engineers for 9/11 Truth | ae911truth.org | @AE911Truth |
| Electroverse | electroverse.net | @Electroversenet |
| Environmental Working Group | ewg.org | @ewg |
| Eluxe Magazine | eluxemagazine.com | @eluxemagazine |
| Gaia | gaia.com | @YourMotherGaia |
| Behold Israel | beholdisrael.org | @beholdisrael |
| Global Healing | globalhealingcenter.com | @GHChealth |
| Australian National Review | anrnews.com | @anr_news |
| Alliance for Natural Health | anhusa.org | @anhusa |
| AltHealthWorks | AltHealthWorks.com | @AltHealthWORKS |

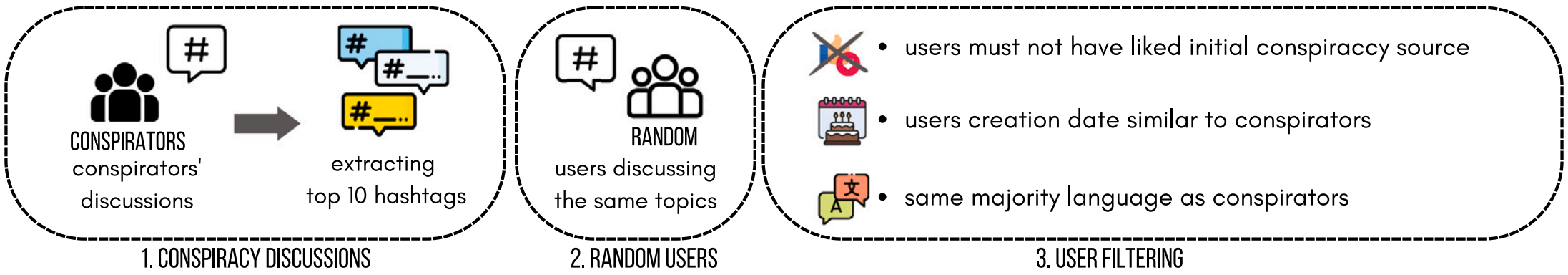

**Fig. 2.** Overview of the proposed strategy for collecting random users.

**Table 3**
Number of users who distributed at least Y likes (up to 35) on at least X distinct S accounts (up to 7).

| Total likes ($l$) | Total sources ($s$) | | | | | | |
|---|---|---|---|---|---|---|---|
| | $s \geq 1$ | $s \geq 2$ | $s \geq 3$ | $s \geq 4$ | $s \geq 5$ | $s \geq 6$ | $s \geq 7$ |
| $l \geq 1$ | 345,936 | 132,828 | 55,064 | 19,966 | 7618 | 2462 | 790 |
| $l \geq 5$ | 147,108 | 95,526 | 49,535 | 19,630 | 7618 | 2462 | 790 |
| $l \geq 10$ | 88,628 | 61,635 | 34,542 | 15,817 | 6932 | 2366 | 779 |
| $l \geq 15$ | 61,778 | 44,206 | 25,228 | 12,118 | 5809 | 2095 | 726 |
| $l \geq 20$ | 46,220 | 33,507 | 19,265 | 9413 | 4807 | 1812 | 648 |
| $l \geq 25$ | 36,055 | 26,342 | 15,169 | 7394 | 3928 | 1517 | 557 |
| $l \geq 30$ | 29,577 | 21,870 | 12,746 | 6370 | 3438 | 1354 | 504 |
| $l \geq 35$ | 24,597 | 18,308 | 10,810 | 5474 | 3000 | 1187 | 447 |

**Table 4**
Top 10 hashtags used by conspiracy users.

| Hashtag | No. of tweets |
|---|---|
| #Covid19 | 31,275 |
| #Bitcoin | 29,496 |
| #Ukraine | 18,524 |
| #NoVaccinePassports | 13,437 |
| #Pfizer | 12,025 |
| #StopTheTreaty | 11,813 |
| #cdnpoli | 10,704 |
| #Canada | 9736 |
| #FreedomConvoy2022 | 9256 |
| #NoVaccinePassportsAnywhere | 9197 |

### 3.1.2. Strategy for collecting random users

To enable a meaningful comparison with our conspiracy group, we establish a set of random users as control group. The selection criteria ensure parity in terms of discussed topics, account creation period, and language usage. This comprehensive approach involves three steps, as shown in Fig. 2:

- **Step 1 — Collecting Topic-Related Discussions**: We collect tweets linked to the top 10 hashtags used by conspiracy users ( Table 4). We establish June 13th, 2022 as the end date for this data collection process. This specific date aligns with the termination of the collection of conspiracy theorists' timeline.
- **Step 2 — Extracting Users discussing this Topics**: This step resulted in the retrieval of 152,588 tweets authored by 82,796 distinct users.
- **Step 3 — Filter Random Users**: We exclude users engaging with any of the 26 conspiracy seed sources and any of the accounts liking those sources, to ensure a non-conspiracy profile. Additionally, we ensure uniformity of the predominant language in tweets. Finally, chose random users whose creation dates match with conspiracy users, maintaining equal distribution among these groups.

This rigorous approach results in a set of 7394 random users. We gather their timelines, providing a comprehensive dataset for comparative analysis, ending up with 19,268,801 tweets.

### 3.2. Feature extraction

We aim to identify the features that separate conspirative users from control users. We use data and meta-data to compute 93 different features that provide insights into various aspects of our users. Each feature is either a continuous numeric value, a binary value, or a set of statistics calculated from distribution (i.e., minimum, maximum, median, mean, standard deviation, skewness, and entropy). Each feature falls into one of three categories: continuous numeric values, binary values, or statistical measures derived from distributions (such as minimum, maximum, median, mean, standard deviation, skewness, and entropy). Some of these features are drawn from prior studies on bots and trolls, specifically selecting those proven effective in detecting or characterizing social bots and state-backed trolls [47].

**Table 5**
The extracted features for each trait [47]. The distribution parameters are the *min*, *max*, *mean*, *median*, *std*, *skewness*, and *entropy*.

| Feature | Type | Description |
|---|---|---|
| *Credibility* | | |
| Following Count | Numeric | Number of followings |
| Followers Count | Numeric | Number of followers |
| Followers Ratio | Numeric | Ratio between the number of following users and the number of followers squared |
| Account Age | Numeric | Account age expressed in days |
| Followers Ratio | Numeric | Ratio between followers and age |
| Following Ratio | Numeric | Ratio between followings and age |
| Tweets Ratio | Numeric | Ratio between the number of tweets and age |
| Verified | Binary | If the account is verified or not |
| Has bio | Binary | If the account has got the bio |
| Has Default Pic | Binary | If the account has got the Twitter default pic or not |
| Has URL in Bio | Binary | If the account has got any URL in the bio |
| URLs Count | Numeric | Number of URLs in the bio |
| Hashtags Count | Numeric | Number of hashtags in the bio |
| Listed Count | Numeric | Number of public lists that the account is a member of |
| Bio Sentences | Numeric | Number of sentences in the bio |
| Bio Tokens | Numeric | Number of tokens in the bio |
| Bio Chars | Numeric | Number of chars in the bio |
| *Initiative* | | |
| Retweet Ratio | Numeric | Ratio between the number of retweets and the number of tweets |
| Reply Ratio | Numeric | Ratio between the number of replies and the number of tweets |
| Tweet-URL Ratio | Numeric | Ratio between the number of tweets containing a URL and the number of tweets |
| Retweet-URL Ratio | Numeric | Ratio between the number of retweets containing a URL and the number of tweets |
| Reply-URL Ratio | Numeric | Ratio between the number of replies containing a URL and the number of tweets |
| Unique Words in Tweets | Distribution parameters | Distribution of the number of unique words in tweets |
| Entropy of Unique Words in Tweets | Distribution parameters | Distribution of the number of unique words entropy in consecutive pairs of tweets |
| *Adaptability* | | |
| Language Novelty | Distribution parameters | Percentage of new tokens in a tweet compared to those previously used |
| Time Between Tweets | Distribution parameters | Distribution of time differences between consecutive tweets |
| Time Between Retweets | Distribution parameters | Distribution of time differences between consecutive retweets |
| Time Between Mentions | Distribution parameters | Distribution of time differences between consecutive tweets containing mentions |
| Retweeted Accounts | Distribution parameters | Distribution of the number of retweeted accounts |
| URL Domains | Distribution parameters | Distribution of the number of domains contained in tweets |
| Tweets Words | Distribution parameters | Distribution of the number of words contained in tweets |
| Tweets Characters | Distribution parameters | Distribution of the number of characters contained in tweets |

In details, our approach involves extracting account features organized into three groups that capture different aspects of social network behavior. These groupings are inspired by previous research [47] that identified attributes related to account trustworthiness, topical focus [51], behavioral dynamics [51,52], and strategic goals [51]. These feature groups, referred to as "traits", offer a suitable framework for describing and distinguishing diverse types of social network accounts. Unlike conventional categorizations (e.g., user-based, friends, network, temporal, content, sentiment, etc.) found in other studies [53–55], our intuitive grouping aligns with the diverse roles an account can assume within the social network context, providing a more nuanced framework for analysis. Table 5 summarizes the features by category. Each feature category serves a specific purpose in capturing distinctive aspects of these users, contributing to the detection and profiling of conspiracy theorists. Credibility features assess the trustworthiness of their profiles, Initiative features capture their active role in shaping discussions, and Adaptability features reveal how they adapt to evolving conditions. The combination of these three feature categories provides a holistic approach to understanding and profiling conspiracy theorists. Together, these features contribute to a nuanced and comprehensive analysis for the effective detection and profiling of conspiracy theorists within the social network. We better discuss them as follows.

### 3.2.1. Credibility features

The rationale behind this feature category is that conspiracy theories often involve discussions that may lack credibility and trustworthiness. Analyzing the credibility features allows us to assess the trustworthiness of social media users based on their profile characteristics. By examining attributes such as the quantity and nature of social relationships, account age, and activity level, these category can help differentiate between low-credibility and high-credibility accounts. This is crucial because discussions involving credible users are more likely to be perceived as organic, providing insights into the authenticity of the content. These features primarily draw from profile metadata, easily observable and assessable when viewing a social network account. These attributes have long served as discriminators for simplistic fake accounts [54,56–58].

### 3.2.2. Initiative features

The rationale behind this group lies in the fact that conspiracy theorists may actively shape online conversations, initiating discussions, and generating content that aligns with their beliefs. Understanding the initiative features allows us to quantify the quality and quantity of a user's activity in shaping discussions. To achieve this, we employ a set of features that quantify the quality and quantity of a user's activity, building on prior works [47,59]. These features include metrics like the ratio of original to retweeted content, indicating an account's contribution to generating fresh and unique material. This goes beyond mere amplification and helps measure the quality and diversity of online discussions, shedding light on the active role played by conspiracy theorists in shaping the narrative. Additionally, metrics like the ratio of tweets to replies reflect the user's engagement in dialogues and exchanges with other users, rather than just broadcasting

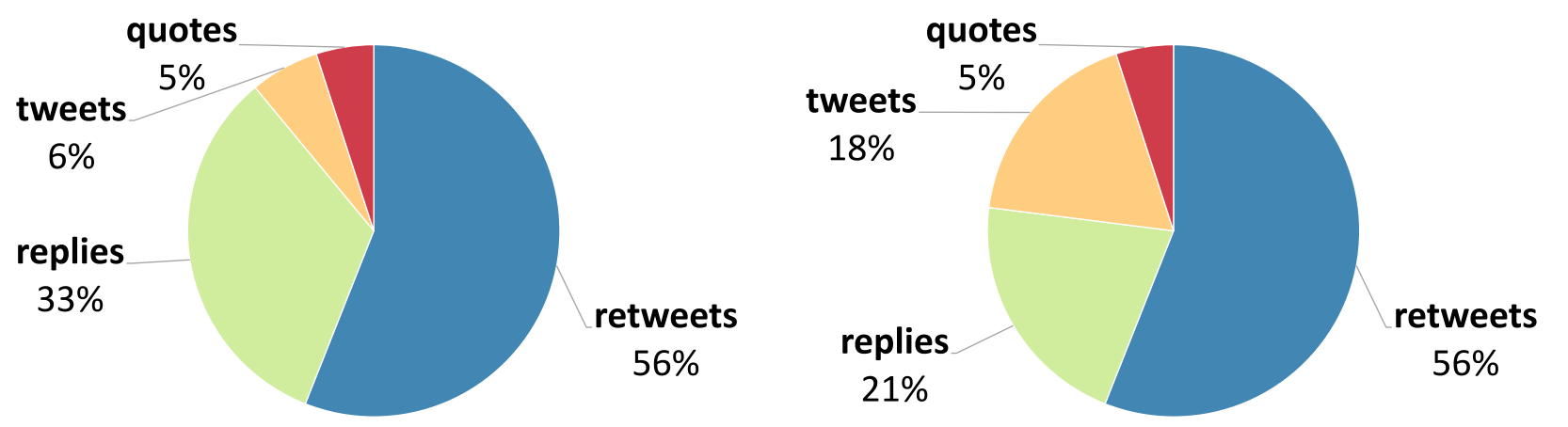

(a) Over 18, 273, 565 total tweets from *conspirative users*

(b) Over 19, 268, 801 total tweets from *random users*

**Fig. 3.** Dataset statistics.

its own messages. These features help measure the quality and diversity of online discussions.

### 3.2.3. *Adaptability features*

The rationale behind this category lies in the fact that conspiracy theorists may adapt their behavior and profile over time in response to emerging or contributed topics. Studying adaptability features allows us to measure the account's willingness to change in order to suit different conditions. By examining linguistic aspects such as language novelty, entropy, and diversity, alongside other temporal changes, adaptability features provide a lens into the evolving nature of conspiracy theorist behavior. This includes how they respond to changing topics and conditions in the social network, offering valuable insights into the dynamic aspects of their engagement. Thus, adaptability ties into the account's temporal and topic-related dynamics and its language usage [45,51,59].

## 4. Results

### 4.1. *Dataset creation (RQ1)*

In literature, users engaged in conspiracy activities are often identified as accounts who employ specific conspiracy-related keywords or share URLs from conspiracy websites [20,21,28–33]. However, some of these users may be bots or trolls attempting to spread panic and skepticism in authorities by pushing alternative explanations for events [60–63]. Misinformed users may inadvertently spreading conspiracy theories [34–36]. Our data collection accounts for this subtle difference between malicious users, misinformed users and conspiracy theorists by adopting a strategy that does not rely on the usage of either keywords or URLs.

Our approach is grounded in the idea that "liking" a post expresses approval or support for its content, which does not necessarily apply to sharing [48]. This implies that users who frequently like posts from a particular account are likely endorsing the themes promoted by that account, and this endorsement is even stronger if the user follows the account. Consequently, a user who consistently likes posts from various conspiracy accounts, while also being a follower of at least one such account, is more likely to believe in conspiracy theories. Furthermore, relying solely on conspiracy keywords or URLs to identify conspiracy users results in capturing only those who actively spread and support specific plots. Our strategy, on the other hand, enables us to identify conspiracy users who may not necessarily propagate the theories they believe in, and who might endorse a range of conspiracy theories if the source accounts are of a generic nature. Here, "generic" refers to Twitter accounts that discuss multiple conspiracy theories concurrently. A similar liking-based strategy was used by authors in [21,32], who identified conspiracy users based on their significant liking activity on conspiracy-related posts. In our approach, we also consider the "follow" relationship from users to conspiracy accounts, providing a stronger validation of their affiliation with the conspiracy realm.

For the control group, we aim to include users that represent the broader social media population while minimizing superficial differences between an average random user and a conspiracy theorist. Utilizing anti-conspiracy keywords or URLs (e.g., science-based) [21,29,30] or general keywords related to broad topics [20,28,31] helps reduce these disparities. However, an additional mechanism is required to ensure similarity between the control and conspiracy groups while maintaining the integrity of both. In our work, instead, we ensure that conspiracy users and random users discuss similar topics, are created around the same time period, and use the same predominant language in tweets. It is important to note that these common topics may not be conspiracy-related. To filter potential conspiracy theorists from the control group, we exclude random users who have liked posts from any of our seed accounts. A similar concept of a control group was used in [22], where users with similar initial activity to conspiracy theorists were identified and tracked as they diverged over time.

We briefly provide an overview of the characteristics of our dataset. Fig. 3 shows descriptive statistics of the collected users. In terms of content, random users exhibit a higher proportion of original tweets (19%) compared to conspirative users. Conversely, the latter group displays a greater inclination towards engaging in replies (35% as opposed to 21%). Regarding retweets and quotes, no substantial differences are evident.

Finally, we verify the presence of automated accounts by employing *Botometer v4* to compute the *bot* scores for both conspiracy and random users [64]. The analysis revealed no bot presence within the conspiracy group. However, approximately 1.5% of random users show a likelihood of being bots with a confidence level greater than 90%. We consider the 1% noise acceptable for our study's purposes.

### 4.2. *Topic characterization (RQ2)*

In this section, we answer to RQ2, by focusing on the key subjects discussed by the two distinct groups. First, we extract the main topics through social network analysis based on co-occurring hashtags. Then, we employ topic modeling to highlight the primary themes of conversation within each user group. In this way, we highlight highly correlated words that may give a hint on the attitude towards a specific topic. Notably, conspiracy users were collected by leveraging likes, while random users were collected by leveraging hashtags over the same time period. Taking that into consideration, we perform the following analysis on timelines and properly handle hashtag seeds, as detailed as follows.

### 4.2.1. *Visualizing co-occurring hashtags per group*

We begin by computing and visualizing the graph of co-occurring hashtags for conspiracy users. We compare it with the graph generated from hashtags that co-occur in tweets posted by random users. These co-occurrence graphs depict the interconnections between hashtags based on their simultaneous appearance within tweets. For clarity, Figs. 4 and 5 show only the top 50 hashtags based on weighted degree.

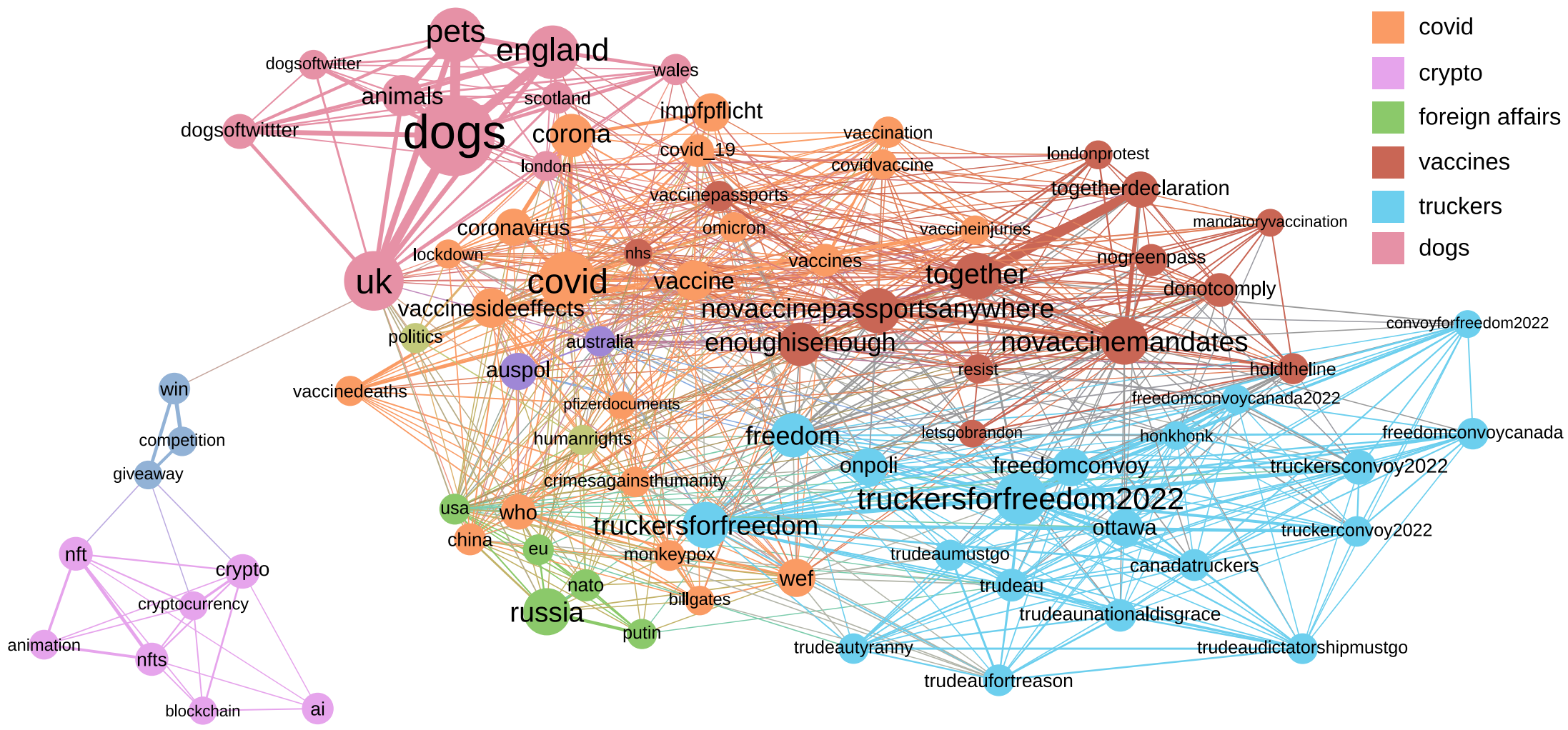

**Fig. 4.** Co-occurrence graph of hashtags mentioned by conspiracy users.

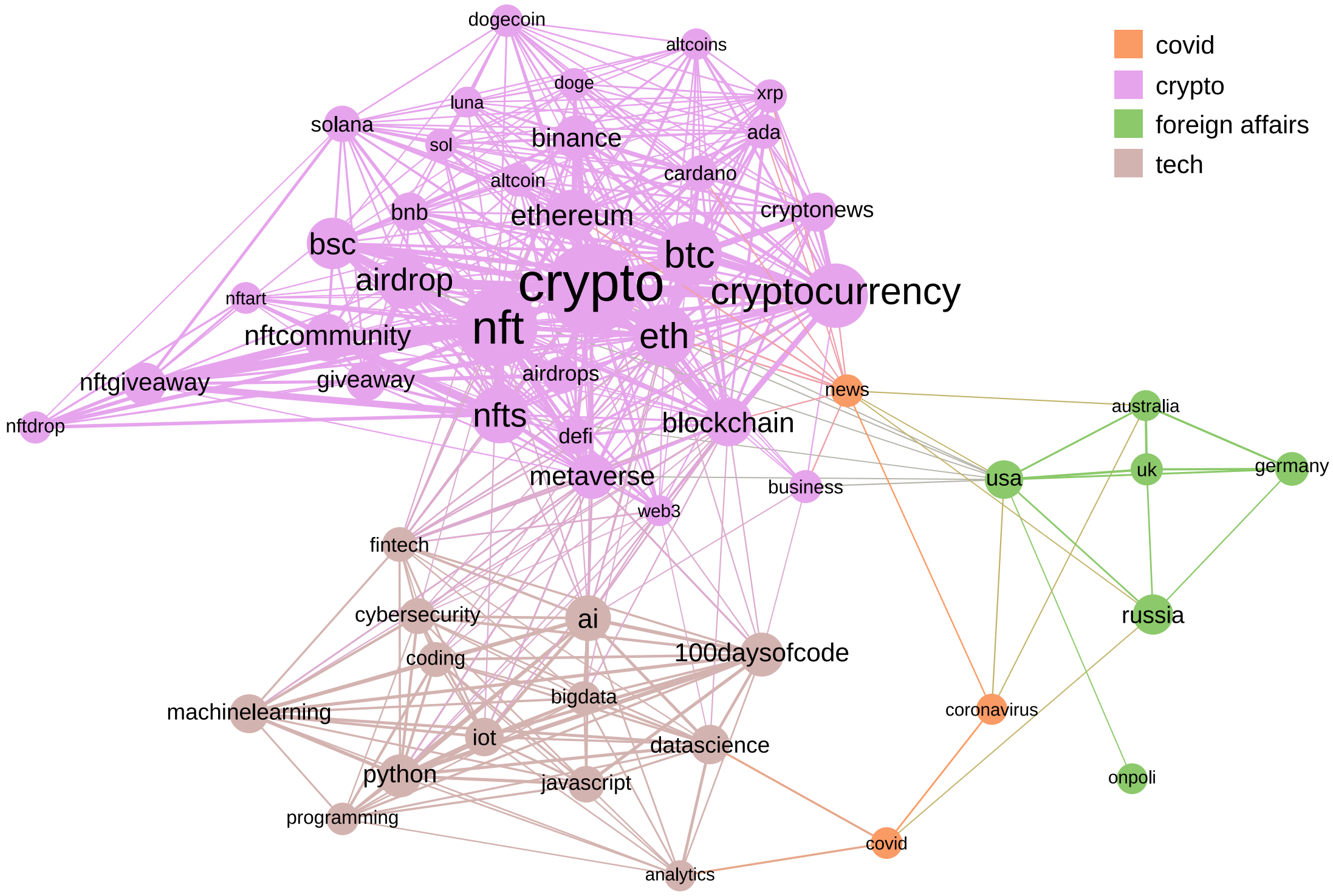

**Fig. 5.** Co-occurrence graph of hashtags mentioned by random users.

Fig. 4 shows the co-occurrence graph of hashtags mentioned in all tweets posted by conspiracy users. As shown, the core of this graph is predominantly composed of two clusters. One cluster centers around topics related to covid-19 and vaccination discourse. The other cluster involves hashtags commonly used for describing images on Instagram [65,66], possibly due to cross-platform social media sharing. In Fig. 5 we reconstruct the co-occurrence graph of hashtags used by the random users. Given that we collected data for these users by focusing on the top 10 hashtags used by conspiracy users, we omit those hashtags from our analysis. In this scenario, the core of the graph is mainly composed of hashtags associated with cryptocurrency. Notably, covid-related hashtags appear on the periphery of the graph. This suggests that during the data collection from random users, cryptocurrency held a stronger influence than the other topics supplied as input. Nevertheless, the popularity of certain topics (e.g., cryptoworld) might surpass others like (e.g., covid), based on factors such as current trends and individual user preferences.

In the next section, we provide a more extensive exploration of the topics and analyze the different user groups' attitude and stances on these subjects.

### 4.2.2. *Characterizing topic discussions and attitudes*

To gain a deeper understanding of the different attitudes towards the online discourse between conspiracy and random users, we employ topic modeling using a recent, advanced, cutting-edge algorithm known as Anchored Correlation Explanation (CorEx) [67]. The CorEx algorithm learns latent topics from documents without assuming an underlying generative model. It maximizes the correlation between groups of words and latent topics, leveraging the dependencies between words in documents. This approach ensures enhanced flexibility, enabling hierarchical and semi-supervised variants [67]. An essential feature of CorEx is also the ability to anchor words, which facilitates semi-supervised topic modeling and enhances topic separability with minimal intervention. Anchoring involves injecting prior knowledge (anchor words) into the topic model to identify and differentiate underrepresented or significant topics. This process enables us to extract pertinent topics and the associated terminology. Given our focus on studying the attitude towards shared main topics by both user groups, we capitalize on the word anchoring capability of CorEx to enhance topic separability.

**Table 6**
Topic modeling results, obtained by applying Anchored Correlation Explanation (CorEx) to conspirative and random users. Conspirative users are characterized by the use of more extreme and intense words when discussing a relevant topic.

| Topic | Highly correlated words by conspirators | Highly correlated words by randoms |
|---|---|---|
| covid19 | billgates, vaccinesideeffects, wakeup, greatreset, vaccinemandate, vaccinedeaths, bigpharma, freespeech | coronavirus, covid19vaccine, fakenews, humanrights, breakingnews, donaldtrump |
| bitcoin | fiat, coins, transactions, decentralized, satoshi | ethereum, cryptocurrency, binance, wallet, nft |
| ukraine | ukrainerussiawar, musk, sanctions, inflation | russians, biden, invasion, war |
| pfizer | adverse, myocarditis, clinical, covid, vaers, pcr | astrazeneca, pcr, schwab, biontech, wef, klaus |
| stopthetreaty | stopthewho, billgatesbioterrorist, trudeaufortreason, crimesagainsthumanity, wefpuppets, reinstatenickhudson, petitions, henchmen, experimental | force, national, service, air, anti, cause, court, population |

We build two distinct models for conspiracy users and random users to account for potential variations in topics and forms of speech. We select the top 10 hashtags from Table 3 as anchor words. After experimenting with various configurations, we set the expected number of topics to 10, as additional topics yielded negligible correlation improvement. Finally, we rank the resulting topics based on the correlation fraction they explain. The outcomes of this analysis are summarized in Table 6, with topics ordered by the amount of total correlation explained. Within each topic, words are arranged according to mutual information with the topic, and anchor words are highlighted in bold. Anchoring substantively augmented the contribution of topics of interest to the model's correlation. High topic quality is confirmed by the presence of non-anchored words with strong coherence within each topic. We report the most informative topics, in Table 6. We uncover some notable differences in the discussion of the same topics. For instance, conspiracy users discussing the topic of covid-19 topic deploy other highly correlated non-anchored words tied to conspiracy terminology [68,69]. Notably, phrases like *wake up* and appeals for *free speech* stand out. Other terms encompass *bigpharma* and *vaccinedeaths*. In contrast, random users use milder language in relation to this topic, such as *fakenews*, *humanrights*, and *breakingnews*, highlighting the moderation of this group. Similarly, words strongly correlated with the "pfizer" topic among conspiracy users revolve around adverse symptoms (e.g., *adverse*, *myocarditis*, *vaers*). Random users, on the other hand, use more general terms (e.g., *biontech*, *wef*).

Another example pertains to the discussion about the international treaty for pandemics prevention and preparedness established by the World Health Organization (WHO) aiming to ensure equitable sharing of vaccines, drugs, and diagnostics during future pandemics [70]. Conspiracy users' correlated words include *billgates bioterrorist*, *trudeaufortreason*, *crimesagainsthumanity*, and other words with a nuance of adversion against the act.[5] In contrast, words used by random users are more generic, and neutrale (e.g., *population*). Finally, regarding discussions on Ukraine and cryptocurrency, no substantial variations emerge between the two user groups.

[5] https://www.reuters.com/article/factcheck-who-treaty-idUSL2N2XH0KA

**Table 7**
Dataset composition (ground-truth and train/test split) for the classification task.

| Class | Users | Split | |
|---|---|---|---|
| | | *Training set* | *Test set* |
| conspiracy | 7394 | 5915 | 1479 |
| random | 7394 | 5915 | 1479 |
| **Total** | 14,788 | 11,830 (80%) | 2958 (20%) |

### 4.3. *Leveraging classification for extracting conspiracy discriminating features (RQ3)*

In this section, we begin by examining key features derived from the bot and troll literature to effectively characterize conspiracy theorists and distinguish them from typical Twitter users. Subsequently, we integrate psycholinguistic traits into our analysis. Finally, we assess the performance of our top detector against a state-of-the-art technique and across a diverse dataset.

In order to identify conspiracy-related users and determine the key features that differentiate them from regular users, we leverage a set of 13 off-the-shelf machine learning algorithms (i.e., Light Gradient Boosting Machine (LIGHTGBM), Random Forest (RF), Gradient Boosting Classifier (GBM), Ada Boost Classifier (ADA), Extra Trees Classifier (ET), Decision Tree Classifier (DT), Logistic Regression (LR), Linear Discriminant Analysis (LDA), Ridge Classifier (RIDGE), K Neighbors Classifier (KNN), Support Vector Machine (SVM), Naive Bayes (NB), Quadratic Discriminant Analysis (QDA)). These classifiers are trained using a stratified 10-fold cross-validation approach. We assess several models, beginning with a baseline model, and progressively adding more features to each subsequent model.

As a preprocessing step, we initially divide our dataset into training and testing sets using an 80/20 split. To provide a rigorous evaluation, we randomly select users for the training and test splits, preserving the balance of the two types of users. As shown in Table 7, the training set includes 12,708 users, 6354% of which are labeled as conspiracy theorists and 6354% as random. The test set consists of 3178 users, of which 1589% are conspiracy theorists and 1589% control users. We address missing categorical values by replacing them with the most frequent value within the respective column. Missing numerical values are substituted with the mean value of their respective columns.

Table 8 shows the outcomes of the classification process. The table presents the performance of two baseline models: *Majority Class*, which always predicts the majority class; and a random predictor. We show alongside the outcomes of the optimal classifier (LIGHTGBM), evaluated in terms of the F1 score [71]. While baseline performance fall short, when applying the LIGHTGBM classifier, we systematically integrate diverse sets of features to evaluate their impact. The optimal outcome is achieved when utilizing all three feature categories, yielding an F1 score of 0.89. Following closely, the second-best performance is observed when employing adaptability features and initiative features independently. These findings suggest that identifying conspiracy theorists with this kind of features can be addressed with good results, although not reaching an exceptional level. In the next section, we analyze the most important features. Following that, in the subsequent section, we conduct experiments by incorporating psycholinguistic features.

**Table 8**
Classification results for the detection of conspiracy theorists for different groups of features. We slightly adjusted the dataset for applying and comparing results using the ConspiDetector tool, by excluding 18 random users due to insufficient text content for computing personality trait features. Best results in each evaluation metric are shown in bold.

| Model | Feature category | Evaluation metrics | | |
|---|---|---|---|---|
| | | *precision* | *recall* | *F1* |
| *Baselines* | | | | |
| Majority Class | – | 0.53 | 1.00 | 0.69 |
| Random | – | 0.49 | 0.49 | 0.49 |
| LightGBM | credibility | 0.79 | 0.77 | 0.78 |
| LightGBM | initiative | 0.79 | 0.85 | 0.82 |
| LightGBM | adaptability | 0.80 | 0.85 | 0.82 |
| LightGBM | credibility + initiative + adaptability | **0.89** | **0.88** | **0.89** |
| *Adjusted dataset for psycholinguistic analysis and comparison* | | | | |
| LightGBM | psycholinguistic | 0.89 | 0.89 | 0.89 |
| ConspiDetector | psycholinguistic | 0.88 | 0.91 | 0.89 |
| LightGBM | credibility + initiative + adaptability + psycholinguistic | **0.94** | **0.94** | **0.94** |

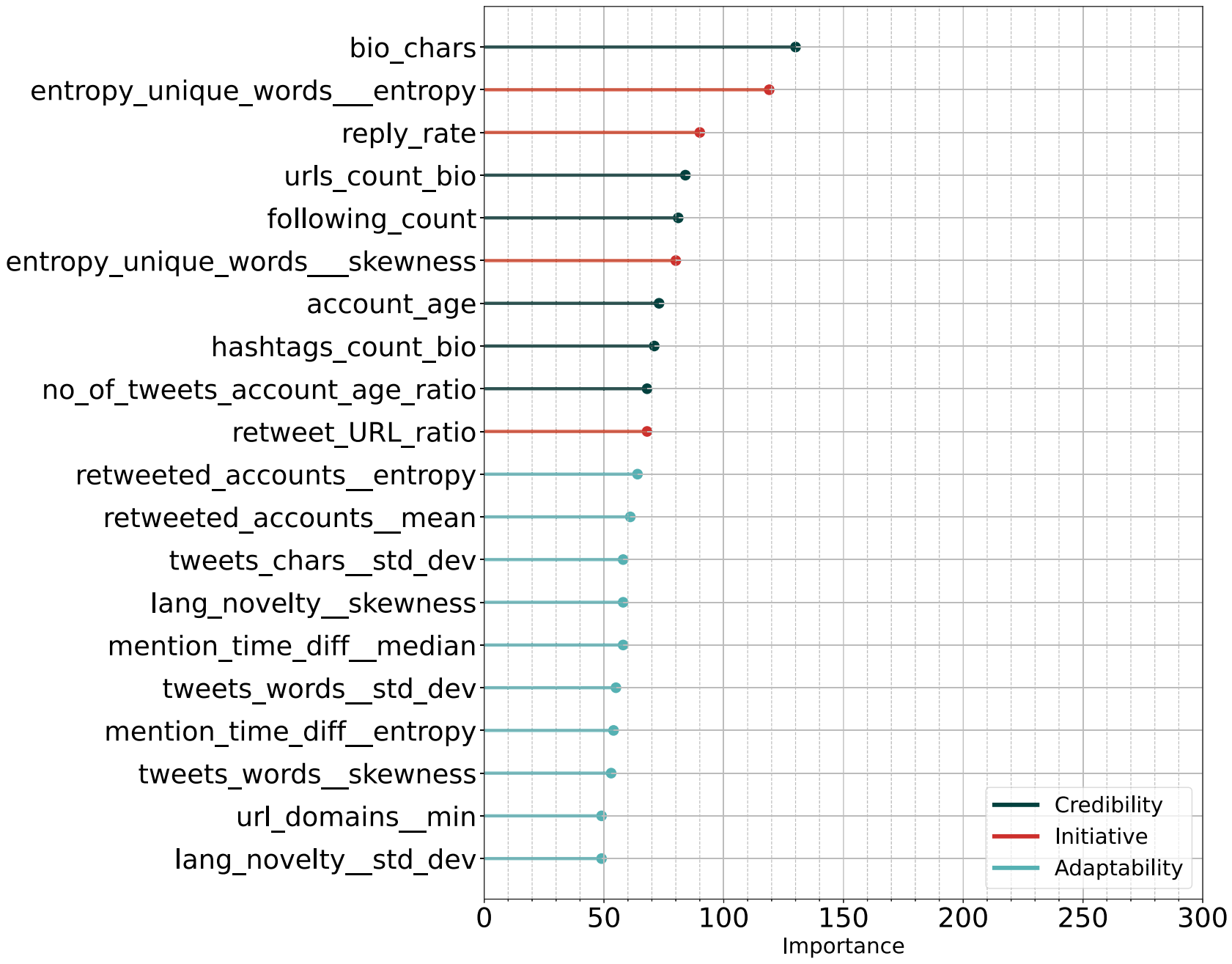

**Fig. 6.** Feature importance considering credibility, initiative and adaptability traits.

*4.3.1. Feature importance evaluation*

Here, we explore the feature importance of the best-performing classifier, specifically the LIGHTGBM algorithm, in a comprehensive model that incorporates all features related to credibility, initiative, and adaptability. The goal is to gain a better understanding of which features contribute significantly to the accurate identification of users engaging in conspiracy activities.

Fig. 6 exhibits the features in descending order of their impact on the Gini criterion, providing insights into their predictive importance within the model. The figure highlights the top 20 features, offering a ranked view of their significance in predicting conspiracy-related users. Feature importance plays a crucial role in comprehending the dynamics of various phenomena where several key factors emerge.

Among the discriminating features, the length of the biography emerges as the most influential. A closer examination, as demonstrated in Fig. 7(a), reveals that random users tend to have longer bios, while some conspiracy theorists notably lack a bio altogether. Interestingly, all random users consistently have bios. This discrepancy may suggest a divergence in communication styles and online presence strategies between the two groups. The prevalence of bios among random users could indicate a conscious effort to provide more comprehensive self-descriptions, potentially aligning with a diverse range of interests or affiliations. On the other hand, conspiracy theorists opt for shorter or absent bios in Twitter profiles, possibly driven by a preference for online anonymity, a focus on content sharing over personal details, and a skepticism towards mainstream identity norms. This choice reflects a distinct approach to self-representation aligned with the perceived values and practices within the conspiracy theorist community.

As the second most discriminating feature, the entropy of the number of unique words in consecutive pairs of tweets offers deeper insights into the linguistic patterns and communication styles of users. Although median differences were not observable, the pronounced skewness towards higher entropy values in the distribution for conspiracy theorists indicates a wider range and variability in the vocabulary they employ across successive tweets, as shown in Fig. 7(b). While conspiracy theorists may often use recurring expressions (e.g., “Wake up!”, “Question

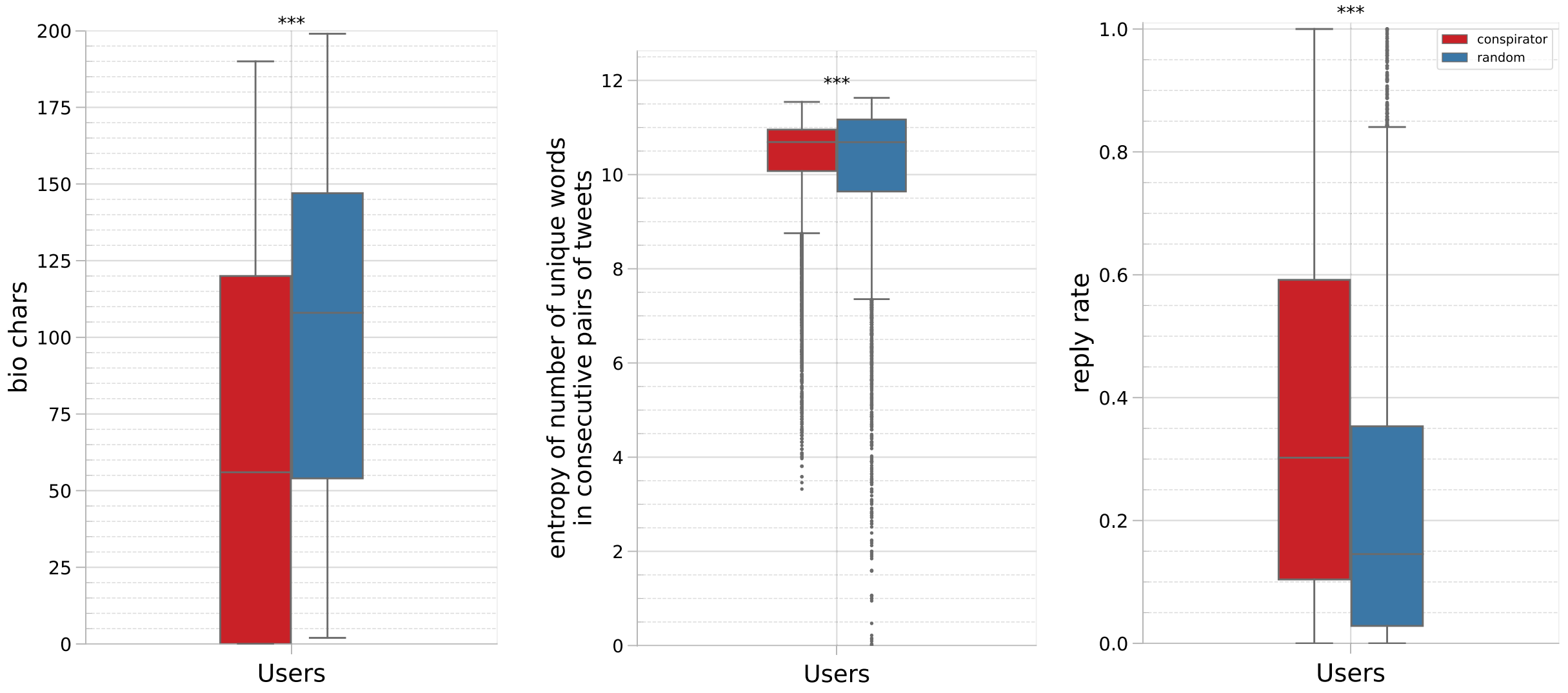

(a) Random users write longer biographies than conspiracy theorists.

(b) While there are no median differences in gauging the unpredictability of unique word occurrence in consecutive tweet pairs, the distribution seems more skewed among conspiracy theorists compared to random users.

(c) Conspiracy theorists are characterized by a higher reply rate with respect to random users.

**Fig. 7.** Most discriminating features of the model.

everything!”), the elevated entropy may indicate diverse linguistic patterns stemming from the broad spectrum of topics covered and individual variations in expression styles. This variability could also be influenced by adaptability to current events and external discourse, contributing to a dynamic and unpredictable use of language within the conspiracy theorist community.

As third discriminating feature, the reply rate provides insights into the level of engagement a tweet generates and its consequential relevance and impact on the audience. A high reply rate implies a user’s interest in participating in discussions and encourage interactions. Fig. 7(c) reveals that conspiracy theorists exhibit a higher reply rate compared to random users. When replying, conspiracy users connect with a wider audience and engage in prolonged conversations with respect to random users.

In addition to the aforementioned features, additional variables contribute to understanding the significance of certain characteristics in the analysis. For instance, the number of shared URLs in the biography (*urls_count*), prevalent among conspiracy theorists, offers insights into the extent of their engagement with external content supporting their beliefs, potentially influencing the reception of their tweets.

In summary, initiative and credibility-related features are the most influential factors in user categorization, followed by adaptability-related features. The credibility features, especially the ones related to the biography (*bio_chars*, *urls_count_bio*, *following_count*, *hashtags_count_bio*) are optimal for discerning conspiracy theorists from random social media users. However, linguistic aspects in tweets are not less important, as the entropy of the number of unique words in consecutive pairs of tweets shows.

In the subsequent section, we compare our findings and feature importance with a cutting-edge technique from the state-of-the-art that primarily focuses on analyzing the psycholinguistic properties of users. This exploration aims to provide deeper insights and understanding of conspiracy activity.

#### *4.3.2. Psycholinguistic features and comparison with a state-of-the-art method*

Here, we conduct a comparison with a methodology from the state-of-the-art [29] that relies on the psycholinguistic traits for identifying conspiracy users and anti-conspiracy users. Similarly to this approach, we additionally leverage:

- *Emotions*: the amount of emotions expressed by the users in their tweets, which includes eight emotional categories (i.e., anger, anticipation, disgust, fear, joy, sadness, surprise, and trust) as defined in [72], computed by leveraging the National Research Council (NRC) emotions lexicon [73].
- *Sentiment*: the amount of sentiment polarity (i.e. positive, negative) expressed by the users in their tweets, computed by leveraging the National Research Council (NRC) sentiment lexicon [73].
- *Personality traits*: we infer the personality traits of users from their tweets by utilizing the IBM Personality Insights API.[6] These traits consist of the renowned *Big Five traits* [74] (agreeableness, conscientiousness, emotional range (or neuroticism), extroversion and openness), five *Values* (conservation, hedonism, openness to change, self-enhancement and self-transcendence) and 12 *Needs* (challenge, closeness, curiosity, excitement, harmony, ideal, liberty, love, practicality, self-expression, stability and structure).
- *Linguistic patterns*: we assess the variety of linguistic patterns exhibited in a user’s tweets using the LIWC tool [75]. In particular, we extract pronouns (I, we, you, she or he, they), time focus (past, present, future), personal concerns (work, leisure, home, money, religion, death), informal language (swear, assent, non-fluencies, fillers), cognitive processes (causation, discrepancy, tentative, certainty) and affective processes (anxiety).

[6] https://personality-insights-demo.ng.bluemix.net/ As of December 2021, this API is no longer supported or maintained. However, requests to the demo API (https://personality-insights-demo.ng.bluemix.net/) can still be leveraged.

**Table 9**
Dataset composition (ground-truth and train/test split) for *ConspiDetector* [29].

| Class | Users | Split | | |
|---|---|---|---|---|
| | | *Training set* | *Validation set* | *Test set* |
| conspiracy | 7394 | 4462 | 1823 | 1109 |
| random | 7376 | 4451 | 1818 | 1107 |
| **Total** | 14,770 | 8913 (60%) | 3641 (25%) | 2216 (15%) |

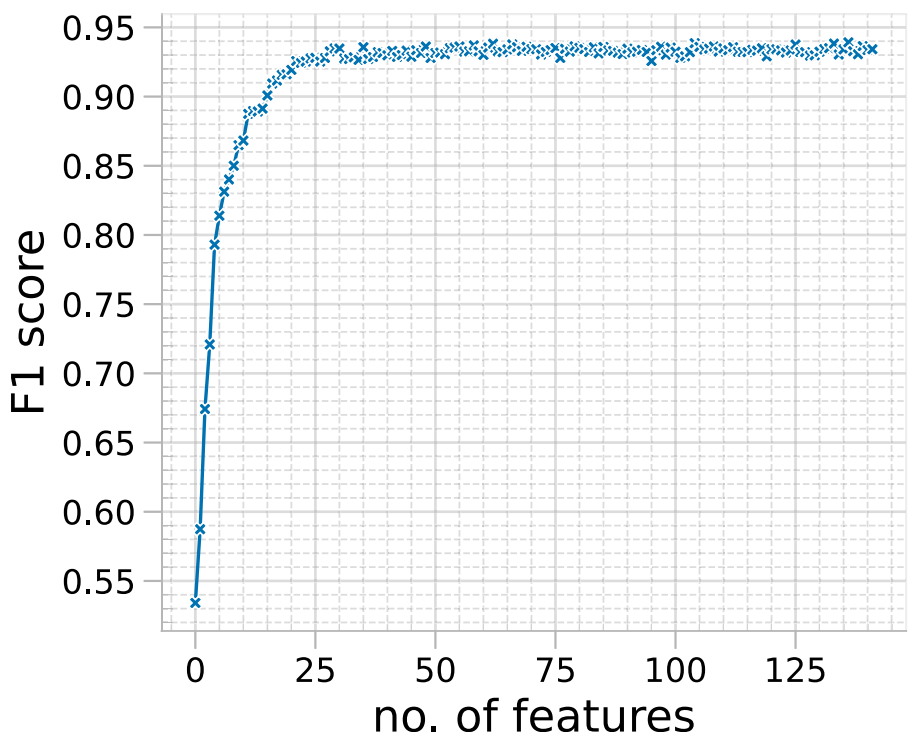

**Fig. 8.** Growth of F1 score with increasing number of features, organized by importance.

Following the methodology called *ConspiDetector* in [29], we incorporate these user-specific characteristics and GloVe embeddings of user tweets in a dual-branch Neural Network. We exclude 18 random users from the analysis due to insufficient text content for computing IBM Personality traits. Table 9 shows the details of the training, validation, and test sets.

First, we assess the performance of *ConspiDetector*, which relies exclusively on psycholinguistic features, on our dataset. We compare this evaluation to the results of our most effective machine-learning classifier model, trained solely on psycholinguistic features. The outcomes are presented in Table 8. Notably, the standard machine learning algorithm trained on our dataset using psycholinguistic features alone yields results similar to those of *ConspiDetector*. Both models achieve an F1 score of 0.89, with *ConspiDetector* exhibiting slightly superior recall performance. Interestingly, a comparable score was also observed in [29] on their dataset. This reflects the overall good performance of psycholinguistic traits individually.

Finally, alongside the evaluation of psycholinguistic features, we re-evaluate our top-performing machine-learning classifier, incorporating these psycholinguist traits along with our credibility, initiative, and adaptability features. The results presented in Table 8 indicate a significant enhancement in our detection capabilities, with an F1 score of 0.94. This outcome suggests that an increase in detectability is achievable by considering social network properties in conjunction with psycholinguistic traits.

Fig. 8 shows the robustness of our conclusions by illustrating the growth of F1 as the number of features increases, arranged by their importance. The Fig. 8 shows that the first most important feature independently achieves a 0.5 F1 score, while a substantial improvement to 0.94 is realized when incorporating the top 30 features. This result suggests that the addressed problem is not quite straightforward or overly simplistic, emphasizing that individual features alone do not reach elevated levels of accuracy.

When looking at the most discriminating features when also considering the psycholinguistic category, Fig. 9 illustrates that psycholinguistic traits are not as pivotal as the main credibility and initiative characteristics previously discussed. Nevertheless, they hold prominence within the top 20 features. Fig. 10 shows that, within the most important psycholinguistics features, the *need for stability*, *need for curiosity* and the *usage of offensive language* are noteworthy. In particular, conspiracy theorists may exhibit a *lesser need for stability* in comparison to a standard social media user due to several psychological factors. Firstly, their inherent openness to unconventional ideas and alternative perspectives fosters a willingness to explore and accept information outside mainstream narratives. This tendency for cognitive flexibility may lead them to *prioritize* intellectual *curiosity* over a need for stability in beliefs. Additionally, conspiracy theorists often harbor deep scepticism and mistrust towards official sources, creating an environment where cognitive dissonance and uncertainty are more tolerated. The desire for uniqueness and nonconformity within the conspiracy theorist community may further contribute to a reduced emphasis on stability, as individuals seek to distinguish themselves by adopting unconventional viewpoints.

Last but not least, the result on the *informal swear* trait suggests that conspiracy theorists use more curse words and offensive language than a standard social media user.

### 4.3.3. Comparison with a state-of-the-art dataset

Finally, to evaluate the robustness and generalizability of our approach, we conduct additional experiments using a different Twitter dataset. While there are other publicly available Twitter datasets focusing on specific conspiracy theories, our primary emphasis is on detecting conspiracy theorists rather than specific conspiracy theories. Consequently, the existing literature offers limited Twitter datasets that align with this approach, as shown in Table 1. Notably, we retrieved the dataset utilized in [28]. This dataset provides a good comparison, as it is focused on the user and collected by leveraging conspiracy hashtags, and comprises 109 conspiracy theorists and an equal number of control users. Each group collectively contributes approximately 300k tweets from the users' timelines, resulting in a dataset comprising over 600k tweets in total. For this experiment, we needed to excluded 50 features from consideration, out of the total 147 features across all four categories: credibility, initiative, adaptive, and psycholinguistic. This exclusion was due to insufficient information in the additional dataset (e.g., missing data include tweet creation date details (such as hours and seconds), absent URLs, etc.).

Following this initial processing step, we perform the experiments on both our dataset and the dataset from [28]. This involves employing our top-performing machine learning algorithm, `LightGBM` with an evaluation carried out using the remaining 97 features.

The outcomes are presented in Table 10. Examining our dataset, the F1 score achieves 0.92, a value slightly lower but comparable to the one obtained when utilizing all 147 features ( Table 8). In the case of the new dataset, employing the same 97 features yields a total score of 0.88. This result reflects well the robustness and generalizability of this approach.

Finally, we present the results obtained by merging two datasets. Specifically, the dataset referenced in [28] consists of 109 users per group, whereas ours encompasses approximately 7K users per group. To ensure balance, we randomly selected 109 users for each group in our dataset. This strategy yielded a well-balanced dataset, comprising 218 conspiracy theorists and 218 random users. The performance using this configuration is presented in Table 10. The model demonstrates effective generalization and achieves an overall good F1 score of 0.87.

## 5. Conclusions

Online conspiracy detection is a challenging task that requires a combination of robust data and tools. In this paper, we proposed a comprehensive methodology for collecting a rigorous Twitter dataset to study conspiracy theorists' characteristics and compare them to randomly selected accounts that exhibit similar characteristics. In particular, we leveraged the "like" behavior on social media platforms as it can reveal affiliation with conspiracy theories better than other behaviors (e.g., retweets, relying on the use of URLs, etc.). In fact, users

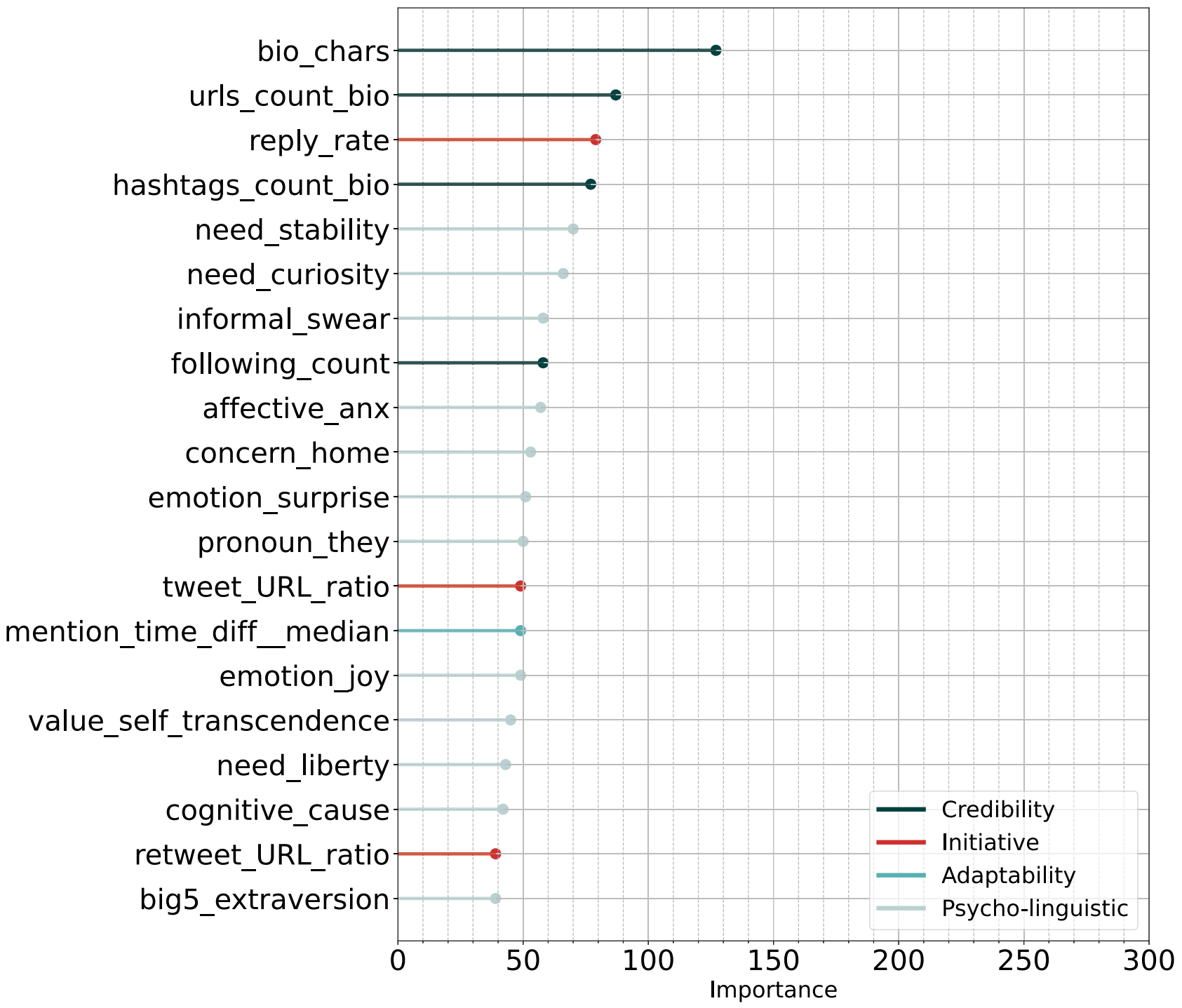

**Fig. 9.** Feature importance considering credibility, initiative, adaptability and psycholinguistics traits.

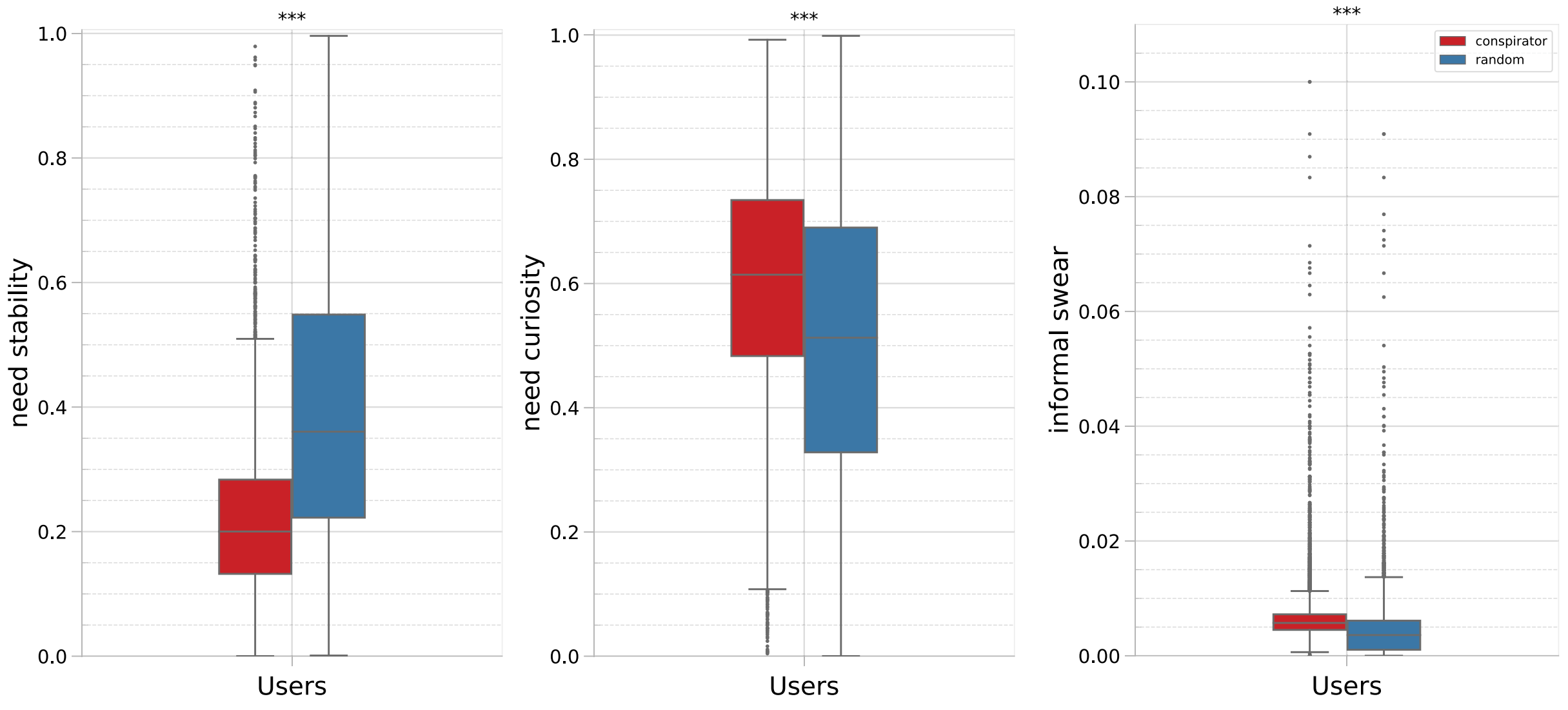

(a) *Stability* trait as defined by IBM Personality Insights API. Conspiracy theorists prioritize a lesser sense of predictability and steadiness in their lives.

(b) *Curiosity* trait as defined by IBM Personality Insights API. Conspiracy theorists are more curious.

(c) *Informal Swear* as defined by LIWC library. Conspiracy theorists use stronger and more offensive language.

**Fig. 10.** Most discriminating psycholinguistic features.

**Table 10**
Comparison with a dataset from the literature.

| Model | Dataset | Feature category | Evaluation metrics | | |
|---|---|---|---|---|---|
| | | | *precision* | *recall* | *F1* |
| *Adjusted features for dataset comparison* | | | | | |
| LightGBM | dataset in [28] | credibility + initiative + adaptability + psycholinguistic | 0.88 | 0.88 | 0.88 |
| LightGBM | our dataset | credibility + initiative + adaptability + psycholinguistic | 0.92 | 0.92 | 0.92 |
| LightGBM | our dataset + dataset in [28] | credibility + initiative + adaptability + psycholinguistic | 0.87 | 0.87 | 0.87 |

who frequently like posts from a specific account are likely to support the themes promoted by that account, especially if they also follow the account. This endorsement of themes is stronger when users both like posts and follow conspiracy-related accounts, making them more prone to believing in conspiracy theories. For the control group, we collected users representing the broader social media population whose activity matches the topics discussed and the account creation time of conspiracy users. In this way, we created a more balanced comparison between conspiracy users and regular users while maintaining the integrity of both groups.

In addition, we presented a robust approach to detect online conspiracy theorists based on their behavioral characteristics, linguistic features, temporal patterns, and other features proposed in the literature for identifying bots and trolls. The goal of this classification task is twofold. On the one hand, we showed that using a standard machine learning classifier on linguistic features and temporal patterns outperforms several baselines and a model proposed in the state-of-the-art as measured by accuracy and F1 score. On the other hand, we employ these findings to profile the two user groups and highlight features that differentially characterize conspiracy-oriented users on social media.

While an account's credibility and proactive initiative engagement are important factors, the findings highlight the significance of linguistic attributes in the analysis. In particular, the identification of conspiracy accounts is augmented by an understanding of psycholinguistic traits. Therefore, it is essential to develop methods that focus on identifying and monitoring conspiracy users based on linguistic traits and patterns, rather than the specific content of their claims. This approach can be pivotal in addressing the spread of conspiracy beliefs on various platforms and domains.

### 5.1. Limitations and future work

Our methodology enables the construction of a comprehensive dataset encompassing a diverse range of conspiracy theorists engaging in a wide variety of conspiracy theories. This section addresses the challenges and limitations of our work while outlining avenues for future research.

Our current approach primarily focuses on well-known and identifiable conspiracy sources curated by Media Bias Fact Check (MBFC), a widely utilized resource for assessing news sources bias and conspiracy. Despite offering a multitude of sources, domains, and news covering a broad spectrum of conspiracy theories, including emerging or niche narratives that may be of considerable influence, the process of categorizing sources as conspiratorial remains subjective. This subjectivity can be influenced by the personal biases or judgments of the evaluators. The evaluation criteria employed by MBFC may lack universal consensus and exhibit variation among individuals. Moreover, the transparency of MBFC's fact-checking methodology may not be fully disclosed, posing challenges in assessing the accuracy and reliability of their evaluations. Additionally, the reliance on manual annotation by experts raises concerns about scalability and potential biases in the labeling process. Future work should address these limitations to ensure multiple sources and perspectives when assessing the conspiracy level of media sources. A move towards a more automated and unbiased approach to source evaluation is essential.

Specifically, efforts should be focused on developing self-supervised approaches to automate the collection of datasets. A promising avenue for future research involves integrating Generative Adversarial Learning (GAN). In the ever-evolving landscape of conspiracy theories, GANs can establish a closed loop between conspiracy source detection and conspiracy theorist user detection. This shift toward automation not only addresses the limitations associated with manual efforts but could also accelerate the scalability, ensure adaptability, and offer the potential to dynamically adapt to evolving conspiracy narratives, ensuring robustness against emerging trends.

Another limitation is our singular focus on Twitter, potentially overlooking the multifaceted nature of online conspiracy discourse across various media. Recent shifts in the Twitter policies further challenge in terms of research applications. To better understand and tackle online conspiracy activities, future studies should encompass data from multiple platforms. In the same context, a further constraint stems from our data collection process, relying on features derived from users' timelines. In fact, this approach can be computationally intensive and susceptible to data availability issues. For real-time detection, an efficient and robust alternative could involve simpler features based on users' current activities and interactions, that do not depend on the users' history. Indeed, examining how online conspiracy discourse changes over time and responds to external events. By studying the temporal dynamics, we can uncover patterns, trends, and shifts in conspiracy discussions. This approach helps to understand how real-world events influence the evolution of conspiracy narratives online. Exploring these temporal dimensions not only enhances the overall understanding of online conspiracy discourse but also provides opportunities to predict and address the impact of false beliefs more effectively. As such, prioritizing research on the temporal aspects can be a valuable addition to refining and advancing detection methodologies. Future research should also explore other social media sources and social network information to gain additional insights about user credibility and influence, as well as develop self-supervised approaches to automate the dataset collection.

In conclusion, acknowledging these limitations ensures an ongoing commitment to refining and advancing our methodology for identifying conspiracy users in the evolving landscape of conspiracy detection. Our work contributes to the growing field of online misinformation and disinformation research, presenting a valuable dataset and methodology for understanding and combating the propagation of harmful and false beliefs.

## CRediT authorship contribution statement

**Margherita Gambini:** Conceptualization, Investigation, Methodology, Visualization, Writing – original draft, Writing – review & editing. **Serena Tardelli:** Conceptualization, Investigation, Methodology, Visualization, Writing – original draft, Writing – review & editing. **Maurizio Tesconi:** Conceptualization, Supervision, Writing – review & editing.

## Declaration of competing interest

The authors declare that they have no known competing financial interests or personal relationships that could have appeared to influence the work reported in this paper.

## Data availability

Data will be made available on request.

## Acknowledgments

We thank for the support by project SoBigData.it, which receives funding from European Union – NextGenerationEU – National Recovery and Resilience Plan (Piano Nazionale di Ripresa e Resilienza, PNRR) – Project: "SoBigData.it – Strengthening the Italian RI for Social Mining and Big Data Analytics" – Prot. IR0000013 – Avviso n. 3264 del 28/12/2021; This work is also partly supported by project SERICS (PE00000014) under the NRRP MUR program funded by the EU – NGEU.